\documentclass[aps,twocolumn,pra,prl,superscriptaddress,preprintnumbers,amsmath,amssymb,nofootinbib]{revtex4}
\usepackage{graphicx}
\usepackage{color}
\usepackage{braket}
\usepackage{comment}
\usepackage{hyperref}

\pdfoutput=1

\newcommand{\be}{\begin{equation}}
\newcommand{\ee}{\end{equation}}
\newcommand{\bea}{\begin{eqnarray}}
\newcommand{\eea}{\end{eqnarray}}

\newcommand{\tr}{\mathrm{Tr}}

\begin{document}
\preprint{OU-HET-970}

\title{Thoughts on Holographic Complexity and its Basis-dependence
}

\author{Koji Hashimoto}\email[]{koji@phys.sci.osaka-u.ac.jp} 
\author{Norihiro Iizuka}\email[]{iizuka@phys.sci.osaka-u.ac.jp} 
\author{Sotaro Sugishita}\email[]{sugishita@het.phys.sci.osaka-u.ac.jp}

\affiliation{Department of Physics, Osaka University, Toyonaka, Osaka 560-0043, JAPAN}

\date{\today}

\begin{abstract}
In this paper, we argue that holographic complexity should be a basis-dependent quantity. Computational complexity of a state is defined as a {\it minimum} number of gates required to obtain that state from the reference state. Due to this minimality,  it satisfies the triangle inequality, and can be regarded as a (discrete version of) distance in the Hilbert space. However, we show a no-go theorem that any basis-independent distance cannot reproduce the behavior of the holographic complexity. Therefore, if holographic complexity is dual to a distance in the Hilbert space, it should be basis-dependent, \textit{i.e.}, it is not invariant under a change of the basis of the Hilbert space. 
\end{abstract}

\maketitle

\setcounter{footnote}{0}

\centerline{\bf {Introduction}}
\vspace{1mm}

What is the boundary dual of the black hole interior in AdS/CFT? 
If such a quantity exists, it should enable us to probe behind the black hole horizon 
in terms of the boundary theory. Therefore it is very interesting to study any candidates for that. 
In particular since the interior of a black hole grows linearly in time for a very long period $t \sim e^S$ where $S$ is the entropy of the system \cite{Susskind:2014rva, Susskind:2016tae} 
(see also \cite{Hartman:2013qma} for earlier observation of the late time behavior of the entanglement entropy), 
the dual quantity should also satisfy this property. 
In fact, it is conjectured that the dual quantity is quantum computational complexity 
\cite{Susskind:2014rva, Susskind:2016tae}, 
since the maximal complexity for a quantum system seems as big as $e^S$, and also complexity, especially that of quantum circuit models, grows linearly in time \cite{Hayden:2007cs, Susskind:2014rva}. 
More concretely, there are two conjectures according to these expectations: the complexity = volume (CV) conjecture \cite{Susskind:2014rva, Stanford:2014jda} and the complexity = action (CA) conjecture \cite{Brown:2015bva, Brown:2015lvg}. 
For eternal two-sided black holes, the CV conjecture states that the complexity is dual to the volume of the maximal time slice anchored at the two given boundary times. 
The CA conjecture instead proposes that the complexity equals the gravitational action on a region so-called the Wheeler-DeWitt (WDW) patch. 
Although the time-dependence of the maximal volume and the WDW action are quantitatively different (see \cite{Carmi:2017jqz}), they show the same behavior at late times; they grow linearly in time and saturate Lloyd's bound.\footnote{If the growth rate is fractional of the value of Lloyd's bound, one can easily define the holographic complexity by multiplying an overall factor to saturate the bound.} 
Here we call all these `bulk-defined-quantities' {\it holographic} complexity, and denote it by $C_\textrm{hol}$. 
Here the point is that  the late time behavior of holographic complexity is universal,   
even though their behaviors except at late times depend on the details of the definition. 

Recently, there has been many works on the holographic complexity \cite{Susskind:2014jwa, Ben-Ami:2016qex, Couch:2016exn, Chapman:2016hwi, Carmi:2016wjl, Ghodrati:2017roz, Carmi:2017ezk, Couch:2017yil, Moosa:2017yvt, Reynolds:2017jfs, Zhang:2017nth, Swingle:2017zcd, Fu:2018kcp, An:2018xhv, Chen:2018mcc, Chapman:2018dem, Auzzi:2018zdu}. 
In order to verify the conjectures, we need to know the properties of complexity in quantum field theories (QFTs). 
However, currently, even a proper definition of it is beyond our reach.  
Recently there were several proposals for the definition of the complexity for QFTs \cite{Caputa:2017urj, Caputa:2017yrh, Bhattacharyya:2018wym, Hashimoto:2017fga, Jefferson:2017sdb, Chapman:2017rqy, Hackl:2018ptj, Alves:2018qfv, Agon:2018zso},
with which the complexity is evaluated and compared with the holographic counterpart. While each argument 
is concrete, the universal way to define the complexity in QFTs is missing.\footnote{In 2D, there is a nice definition of the complexity by the path-integral optimization procedure \cite{Caputa:2017urj, Caputa:2017yrh, Bhattacharyya:2018wym}. It is interesting to investigate if one can generalize this to higher dimensions.}

Thus, a universal requirement for the complexity in QFTs is called for, and 
in this paper we attempt to characterize the complexity in 
QFTs in regard to the properties of the holographic complexity. 
As we will see soon, complexity is a kind of distance between quantum states. 
In fact, a geometric approach is proposed to define a complexity in \cite{Nielsen:2005, Nielsen:2007}, and the relation to the holographic complexity was discussed in  \cite{Brown:2016wib, Brown:2017jil}
(see also \cite{Jefferson:2017sdb, Chapman:2017rqy, Kim:2017qrq, Khan:2018rzm} for attempts to define complexities for QFTs as geometric distances). 
Since there is a variety of distance measures, we need to discriminate proper ones for our purpose.
In quantum information, popular definitions of the distance are basis-independent. 
We argue that a distance corresponding to the holographic complexity should be a \textit{basis-dependent} quantity, \textit{i.e.}, it is not invariant under a change of the basis of the Hilbert space. 
In other words, we show that any basis-independent distance cannot be dual to the holographic complexity.

\vspace{3mm}
\centerline{\bf {Complexity as a distance}}
\vspace{1mm}

Quantum computational complexity (or gate complexity) is a quantity characterizing the difficulty to construct a state from a given reference state. 
Given a set $\{G_\alpha\}$ of elementary unitary operators called gates $G_\alpha$, the complexity $C(\psi,\psi_0)$ of a state $\ket{\psi}$ is defined as the {\it minimum number of gates needed} to construct $\ket{\psi}$ from a reference state $\ket{\psi_0}$.  
Thus, complexity $C(\psi,\psi_0)$ represents how $\ket{\psi}$ is far from $\ket{\psi_0}$ measured with a given gate set. 
Actually, as we will see as follows, complexity has properties of a distance between two quantum states $\ket{\psi}$ and $\ket{\psi_0}$. 

Let's recall the axioms of distance. 
For a general distance $D(\rho, \sigma)$ between two density matrices $\rho$ and  $\sigma$,\footnote{If the two density matrices are pure states, they can be represented as 
$\rho = \ket{\psi} \bra{\psi}$ and $\sigma = \ket{\psi_0} \bra{\psi_0}$, 
for example.} it satisfies the following axioms: 
\begin{align}
	&D(\rho, \sigma)\geq 0. \label{positivity}\\
	&D(\rho, \sigma) = 0 \;\; \Leftrightarrow \;\; \rho=\sigma. \label{zero}\\ 
	&D(\rho_2, \rho_1) + D(\rho_3, \rho_2) \geq D(\rho_3, \rho_1). \label{tri_ineq}
\end{align}
The last axiom is called the triangle inequality. In general, distances also satisfy the symmetric property:   
\begin{align}
D(\rho, \sigma)=D(\sigma,\rho), \label{reverse} 
\end{align}
but we do not require this property.\footnote{Relative entropy satisfies \eqref{positivity} and \eqref{zero} but not \eqref{tri_ineq} above. This can be checked easily for mixed states. For pure states, if two pure states are the same, then relative entropy becomes zero but for {\it any} different pure states, it diverges. 
}

Complexity $C(\psi,\psi_0)$ shares the same properties \eqref{positivity}-\eqref{tri_ineq}:   
\begin{align}
&C(\psi,\psi_0)\geq 0.
\label{cax1}\\
&C(\psi,\psi_0) = 0 \Leftrightarrow \ket{\psi}=\ket{\psi_0}.
\label{cax2}\\ 
&C(\psi_2,\psi_1) + C(\psi_3, \psi_2) \geq C(\psi_3, \psi_1). \label{comp_triangle}
\end{align}
The first and the second properties are trivial: The number of gates are nonnegative, and if we do not use any gate, the final state is nothing but the initial state and vice versa.  
The triangle inequality eq.~\eqref{comp_triangle} follows from the fact that the complexity $C(\psi, \psi')$ counts the {\it minimum number of gates} to reach $\ket{\psi}$ from $\ket{\psi'}$. 
Thus, although it takes discrete values, gate complexity $C(\psi,\psi_0)$ is a sort of a distance between $\ket{\psi}$ and $\ket{\psi_0}$ in the sense that 
it satisfies the axiom of the distance \eqref{cax1} - \eqref{comp_triangle}.\footnote{
	Complexity can also satisfy the symmetric property $C(\psi,\psi_0) = C(\psi_0,\psi)$
	although we do not require it. 
	Actually, if we choose the gate set so that it has all inverse gates of the gates in the set, the complexity satisfies the symmetric property.} 

In addition, holographic complexity takes continuous values, therefore it is very natural to regard it as a distance shared with the properties of the gate complexity. 
However, under a few assumptions, we will show that there is {\it no} basis-independent distance dual to the holographic complexity, 
and thus conclude that holographic complexity should be a basis-dependent distance.  

\vspace{3mm}
\centerline{\bf {Basis-Independence}}
\vspace{1mm}

All of the physical observables are independent of the choices of the basis of the Hilbert space. 
Therefore, it seems to be natural that the quantum distance dual to holographic complexity is also basis-independent. 
In addition to \eqref{positivity}-\eqref{tri_ineq}, the basis-independence requires that distances satisfy
\begin{align}
D(U\rho U^\dagger, U \sigma U^\dagger)=D(\rho, \sigma), 
\label{base-indep}
\end{align}
where $U$ is any unitary operator on the Hilbert space. 
In fact, well-known quantum distances, \textit{e.g.} the trace distance and the quantum angle\footnote{The trace distance is defined as
$D^\textrm{tr}(\rho,\sigma) \equiv \frac12 \tr \sqrt{(\rho-\sigma)^2} $. 
The quantum angle is defined as $A(\rho,\sigma) \equiv\arccos F(\rho,\sigma)$, where $F(\rho,\sigma)$ is the fidelity 
$F(\rho,\sigma)\equiv \tr \sqrt{\rho^{1/2}\sigma \rho^{1/2}}$. These $D^\textrm{tr}(\rho,\sigma)$ and $A(\rho,\sigma)$ satisfy the axioms of distance \eqref{positivity}-\eqref{tri_ineq} \cite{Nielsen}.  It is obvious from the definitions that they are basis-independent. 
}, satisfy this property \cite{Nielsen}.

However, this requirement of the basis-independence imposes a very strong constraint on complexities as follows.   
Let's consider a distance with the basis independence \eqref{base-indep} between two time-evolved states 
$\ket{\psi(t)}$ and $\ket{\psi(t')}$. 
Since eq.~\eqref{base-indep} holds for any unitary operator $U = e^{- i H \delta t}$, the distance satisfies  
\begin{align}
D(\psi(t),\psi(t')) =D(\psi(t-t'),\psi(0)),  
\end{align}
{\it i.e.}, the distance depends on just the difference of times. 
Thus, if the holographic complexity $C_\textrm{hol}(t)$ is dual to such a distance, the triangle inequality  
\begin{align}
&D(\psi(t_1),\psi(0))+D(\psi(t_2),\psi(t_1))\geq D(\psi(t_2),\psi(0))
\end{align}
leads to the following inequality 
\begin{align}
C_\textrm{hol}(t_1) + C_\textrm{hol}(t_2-t_1) \geq C_\textrm{hol}(t_2),  
\label{time_ineq}
\end{align}
where $t_2\geq t_1$. 
We will see that this inequality does not match the known properties of the holographic complexity.  
Therefore, we should reject the requirement \eqref{base-indep}.

\vspace{3mm}
\centerline{\bf {Lloyd bound and holographic complexity}}
\vspace{1mm}

A characteristic property of the holographic complexity is the saturation of the Lloyd bound \cite{Lloyd2000}. 
Lloyd argued that the rate of computation by a physical device is limited by the energy; it is essentially due to the 
uncertainty principle 
\begin{align}
\frac{1}{\Delta t} \sim M .
\end{align} 
Based on this argument, it is conjectured that the growth rate of the holographic complexity is bounded by the mass $M$ of the black hole \cite{Brown:2015bva, Brown:2015lvg}:  
\begin{align}
\frac{d C_\textrm{hol}}{d t}\leq \alpha M, 
\label{comp_bound}
\end{align}
where $\alpha$ is a numerical constant and its value is not important in the following discussion. 

In \cite{Brown:2015bva, Brown:2015lvg}, it is also conjectured that the uncharged black holes saturate the bound \eqref{comp_bound} at late times 
\begin{align}
\lim_{t \to \infty} \frac{d C_\textrm{hol}}{dt} = \alpha M .
\label{late_time}
\end{align}
This is natural in the view that black holes are the fastest scrambler \cite{Sekino:2008he, Maldacena:2015waa}; 
as far as complexity is associated with the growing black hole interiors at late times, 
its evolution is expected to take the maximum speed. 

For example, in the CV conjecture with the definition of the holographic complexity as 
$C_\textrm{hol}(t) = {\mathcal{V}(t)}/{G_N L}$, 
where $\mathcal{V}(t)$ is the maximal volume at time $t$, and $G_N$ is Newton's constant and $L$ is the AdS radius, $C_\textrm{hol}(t)$ for an uncharged planar black hole satisfies 
\begin{align}
\frac{d C_\textrm{hol}}{dt} \leq \frac{16 \pi}{d-1} M, 
\end{align}
for any $t$ \cite{Carmi:2017jqz},\footnote{The time $t$ is different from $t$ in \cite{Carmi:2017jqz} by a factor of two.}   
and saturates the bound at late times \cite{Susskind:2014rva, Stanford:2014jda}. 

On the other hand, in the CA conjecture, the bound \eqref{comp_bound} is violated although it behaves as \eqref{late_time} at late times \cite{Carmi:2017jqz}. See also \cite{Cottrell:2017ayj} where the violation of the Lloyd bound for the holographic complexity is discussed. 
In any case, we next present the incompatibility between the inequality \eqref{time_ineq} and \eqref{comp_bound}, \eqref{late_time}. 

\vspace{3mm}
\centerline{\bf{A no-go theorem}}  
\vspace{1mm}

We now show a no-go theorem that any basis-independent distance which satisfies the inequality \eqref{time_ineq} cannot be compatible with the Lloyd bound \eqref{comp_bound} and its saturation \eqref{late_time} at late times except for the case that the bound \eqref{comp_bound} is saturated for any time.  
Let us set $t_1=t$ and $t_2=2 t$ in the inequality \eqref{time_ineq}. We then have the inequality 
\begin{align}
\label{triangleforChol}
2 C_\textrm{hol}(t) \geq C_\textrm{hol}(2t). 
\end{align} 
If we define the following function $f(t)$, 
\bea
f(t) \equiv \alpha M  t - C_\textrm{hol}(t) \,, 
\eea
then, from the inequality \eqref{triangleforChol}, it must satisfy 
\bea
\label{triangleforf}
 f(2 t) \geq 2 f(t) \,.
\eea

On the other hand, from the Lloyd bound  \eqref{comp_bound}, $f(t)$  clearly satisfies
\bea
\label{positivederivative}
\frac{df(t)}{dt} \ge 0\,.
\eea
We also have $f(t = 0) = 0$ since 
$C_\textrm{hol}(t = 0) = 0$ from \eqref{cax2}. 
With \eqref{positivederivative}, this implies that $f(t)$ is a nonnegative function; 
\bea\label{positive_f}
f \ge 0  \quad (\mbox{at $t \ge 0$})\,.
\eea
Furthermore, the saturation of the Lloyd bound at late times \eqref{late_time} implies that 
\bea
\label{fapproachconst0}
\lim_{t \to \infty}f(t)= \mbox{const.} \equiv f_0 \, \ge 0 \,. 
\eea
If we ignore the constraint \eqref{triangleforf} which comes from the triangle inequality with basis-independence, 
there are infinite number of functions satisfying \eqref{positivederivative}, \eqref{positive_f} and \eqref{fapproachconst0}.
However, these functions cannot satisfy the inequality \eqref{triangleforf} 
except for the case that the Lloyd bound is saturated at any time, {\it i.e.,} $C_\textrm{hol}(t) = \alpha M t$ exactly for any time $t$ (or equivalently, $f(t)\equiv 0$).  This can be easily seen, since the triangle inequality \eqref{triangleforf} at late times implies 
\bea
\lim_{t \to \infty} 2 f(t) \leq  \lim_{t\to\infty} f(2t) \, \, \Leftrightarrow \, f_0 \le 0 \,.
\eea 
In other words, no function can satisfy \eqref{triangleforf} and \eqref{fapproachconst0} unless $f_0 = 0$ exactly. 
Here, $f_0 = 0$ is equivalent to 
\begin{align}
\frac{d C_\textrm{hol}}{d t}= \alpha M   
\label{dCM}
\end{align}
exactly for {\it any} time $t$ from \eqref{positivederivative} and \eqref{positive_f}. 

On the other hand, 
we point out that there is no {\it known} bulk-defined holographic complexity which satisfies this property 
\eqref{dCM} for any time $t$, see \cite{Carmi:2017jqz}.  

Therefore, we conclude that holographic complexity, satisfying the late time behavior \eqref{late_time}, 
cannot satisfy both the Lloyd bound \eqref{comp_bound} and the nature of the basis-independent distance \eqref{time_ineq}.

\vspace{3mm}
\centerline{\bf{Another argument using dimensional analysis}}  
\vspace{1mm}

Here we provide another argument supporting 
our claim that basis-independent distances cannot be dual to holographic complexities, without relying on the Lloyd bound. 
The basis-independent distance might take the following form
\begin{align}
D(\rho, \sigma) = h (\tr\, g(\rho,\sigma))\,,
\label{our_distance}
\end{align}
where $h$ is an arbitrary function $\mathbb{R} \to \mathbb{R}_{\geq 0}$, and $g$ is an arbitrary map from two density matrices to an operator on the Hilbert space.\footnote{We assume that $g(\rho,\sigma)$ can be expanded as series of $\rho$ and $\sigma$. 
}
We concentrate on this class of distances in this section, although we are not sure if all basis-independent distances take the form \eqref{our_distance}.  
For this class of distances, the basis-independence \eqref{base-indep} is automatically satisfied, since they are
defined with the trace.
In addition, the distance is simplified for the case of two pure states $\rho= \ket{\psi}\bra{\psi}$ and $\rho'= \ket{\psi'}\bra{\psi'}$. 
Actually, since pure states satisfy $\rho^2=\rho$, $\tr \rho=1$, and also 
\begin{align}
\tr (\rho \rho')^n = |\braket{\psi| \psi'}|^{2n}\,,
\end{align}
then the distance can {\it always} be written as a function of the fidelity of the two pure states 
\begin{align}
F(\psi, \psi') \equiv |\braket{\psi| \psi'}| \,, 
\end{align}
which satisfies $0 \leq F(\psi, \psi') \leq 1$. (See also \cite{MIyaji:2015mia} for a discussion that a fidelity is related to the bulk volume.)

In literatures, holographic complexities are computed for eternal AdS black holes, which are dual to the thermofield double (TFD) states \cite{Maldacena:2001kr}. 
The TFD state is a pure state 
\begin{align}
\ket{\mbox{TFD}(t=0)} = \frac{1}{\sqrt{Z(\beta)}}  \sum_n e^{-\beta E_n/2} \ket{n}_{\rm L} \ket{n}_{\rm R} 
\label{TFD}
\end{align} 
on the tensor product of the same two Hilbert spaces $\mathcal{H}_{\rm L} \otimes \mathcal{H}_{\rm R}$, 
where 
$Z(\beta)\equiv \sum_n e^{-\beta E_n}$ is the partition function on the single system at the inverse temperature $\beta$.  
Let us consider the time evolution of the TFD state by the total Hamiltonian  
$H_{\rm tot} =H_{\rm L} \otimes {\bf 1}_{\rm R}+{\bf 1}_{\rm L}\otimes H_{\rm R}$, 
and consider the distance between the time-evolved states $\ket{\mbox{TFD}(t)}$ and $\ket{\mbox{TFD}(t=0)}$. 
Since they are pure states, the distance with the form \eqref{our_distance} is a function of their fidelity. 
We represent the fidelity as $F(t)$, which takes a simple expression 
\begin{align}
F(t) \equiv  \, \bigm|\!\braket{\mbox{TFD}(t)|\mbox{TFD}(t=0)}\!\bigm| = \frac{|Z(\beta+2it)|}{Z(\beta)}. 
\label{TFD_fidelity}
\end{align}
Here $Z(\beta+2it)$ means the analytic continuation, $\beta \to \beta +2 it$, of the partition function $Z(\beta)$. 
$|Z(\beta+2it)|^2$ is called the spectral form factor \cite{Dyer:2016pou, Cotler:2016fpe}, whose late time behavior is used to diagnose the discreteness of the black hole spectrum. Interestingly, since the fidelity $F(t)$ is nothing but the normalized spectral form factor, the distance of the TFD state at time $t$ from that at $t=0$ is a function of the spectral form factor. 
We note that the distance function \eqref{our_distance} is introduced {\it independently of states}, so the functions $g$ and $h$ in \eqref{our_distance} should be independent of the temperature. Thus, the temperature- and time-dependence in the distance between the time-evolved TFD states come only thorough the fidelity $F(t)$. 

However, any function of the fidelity $F(t)$ cannot be the candidate for the holographic complexity, which can be seen as follows: 
At high temperature limit for any field theories on $d$-dimensional flat space, 
the leading parts of the partition functions take the following dimensionally determined form,  
\begin{align}
Z(\beta) \sim \exp(c V \beta^{-d})\,,
\label{pf_cft}
\end{align}
where $c$ is a constant which is roughly the number of degrees of freedom, and $V$ denotes the (regularized) spatial volume.\footnote{This formula works even for theories on a curved space, as long as its curvature length scale is much bigger than the inverse of the temperature. }
Using eq.~\eqref{pf_cft}, the fidelity $F(t)$ is given by 
\begin{align}
F(t) =\! \exp\left[cV \!\left(\frac{-1}{\beta^{d}}+\frac{(\beta+2 i t)^d+(\beta-2 i t)^d}{2 (\beta^2+4 t^2)^d}\right)
\right].
\end{align}
At late times $t \gg \beta$, it is written as 
\begin{align}
F(t) \sim \left\{\begin{array}{ll}
\exp \left[ cV \left(\frac{-1}{\beta^{d}}+\frac{(-1)^{d/2}}{(2 t)^d} \right)\right]& (d:\text{even})\\[10pt]
\exp \left[ cV \left(\frac{-1}{\beta^{d}}+\frac{(-1)^{(d-1)/2} d \beta}{(2t)^{d+1}}\right)\right]& (d:\text{even})
\end{array}\right.
\label{late_fidelity}
\end{align}

On the other hand, in the high temperature limit, the mass of the black hole is given by
\begin{align}
M\sim cV/\beta^{d+1}, 
\end{align}  
up to a numerical factor. 
Thus, the late time behavior of the holographic complexity, eq.~\eqref{late_time}, is written as 
\begin{align}
C_\textrm{hol}(t) \sim c V t/ \beta^{d+1} .
\label{Cholbh} 
\end{align} 

Now it is clear that this \eqref{Cholbh} cannot be reproduced by any function of the fidelity \eqref{late_fidelity} without using some other independent function of temperature. 
In other words, 
since the temperature- and time-dependence of any distance comes only through the fidelity, the holographic complexity is not dual to the class of distances \eqref{our_distance}. 
This implies that even though the spectral form factor \eqref{TFD_fidelity} is useful to probe a structure of the spectrum and to diagnose the chaos \cite{Dyer:2016pou, Cotler:2016fpe}, it fails to capture the 
fine-grained structure of the Hilbert space, necessary for complexity.

\vspace{3mm}
\centerline{\bf{Discussions}} 
\vspace{1mm}

We have assumed that complexity can be defined as a distance in the Hilbert space like Nielsen's approach \cite{Nielsen:2005}. 
Then, if we further assume that the distance is basis-independent, we have faced the mismatch with the holographic complexity. 
The assumption that complexity is a distance is probably reasonable because complexity should satisfy the same axioms of distance.  
The mismatch clearly comes from our assumption that the distance is basis-independent \eqref{base-indep}.

Actually, the basis-independence restricts the distances to the class as \eqref{our_distance}. For pure states $\ket{\psi}$ and $\ket{\psi'}$, such a  distance depends only on the absolute value of the inner product $|\braket{\psi|\psi'}|$.  
As explained in \cite{Brown:2017jil}, the inner product $|\braket{\psi|\psi'}|$ loses much of the information of states; any states orthogonal to $\ket{\psi}$ are regarded as the most distant state from $\ket{\psi}$. 
Therefore, we cannot see the fine-grained structure of the system by this class of distances\footnote{Javier Magan and Henry Maxfield pointed out to us that basis-independent distances saturate at $t = O(\beta)$, therefore they cannot be a candidate of holographic complexity. We thank them for discussions of related issues.}. 
However holographic complexity needs to capture the fine-grained structure such that it keeps growing at late time.

Without the basis-independence, we could not obtain the strong constraint from the triangle inequality \eqref{comp_triangle}. 
If we set $\psi_1= \psi(0)$, $\psi_2= \psi(t)$ and $\psi_3= \psi(t+dt)$ in \eqref{comp_triangle}, we obtain 
\begin{align}
C(t,0)+C(t+dt,t)\geq C(t+dt,0)\,,
\end{align}
which leads to the following inequality
\begin{align}
\frac{d C(t,0)}{dt} \leq \frac{d C(t',t)}{dt'}\!\biggm|_{t'=t} \,.
\end{align} 
The inequality is a kind of the Lloyd bound. 
The growth rate of the complexity of $\psi(t)$ with the reference state $\psi(0)$ is bounded from the growth rate between two near states. 
Since this inequality holds between two different reference states, it doe not give a constraint to time dependence of complexities with the fixed reference state.

Our conclusion is that if holographic complexity is dual to a distance in the Hilbert space, it should be basis-dependent. Since all physical observables in quantum mechanics are defined basis-independently,  this might sound a bit puzzling in the following sense; If we admit that holographic complexity is basis-dependent, the growing volume of the wormhole cannot be observable. However this is not a contradiction. To see this, remember that in holographic entanglement entropy \cite{Ryu:2006bv}, RT surface (or volume) is not observable since entanglement entropy is not directly observable.

In this paper, we have considered distances only between states at different times, and did not see the detailed dependence of the reference states. 
In literatures \cite{Jefferson:2017sdb, Chapman:2017rqy},  
unentangled states (or direct products states) are often taken as reference states. 
Here, ``unentangle" means forming a direct product structure under the spatial (or geometrical) decomposition of the Hilbert space, and thus a specific basis respecting the spatial structure or locality is implicitly chosen.  
On the other hand, the basis-independent distances do not respect locality at all. 
Since basis-independence Eq.~\eqref{base-indep} was required for arbitrary unitary transformations including highly nonlocal ones, this turns out too strong assumption. 
To seek for a good definition of complexity as a distance, we should respect the locality of quantum field theories. 
Representation of states by tensor networks, which gives a nice interpretation of growth of the wormhole \cite{Hartman:2013qma},  also suggests that we should respect the locality. If we perform the non-local transformations, such a local tensor structure of the state is lost.   
Furthermore, in order to define complexity, one needs to define, at first, a gate-set. If one changes the basis, the gate-set also changes and accordingly, the complexity (or distance in the Hilbert space) also changes for fixed target and reference states.\footnote{If one changes the basis  
for states and gates in the same way, then the complexity is kept intact.}
This implies that there should exist a preferred choice of a gate-set which respects the locality of quantum field theories.  In spin systems, to find such a locality-respecting gate-set is not that difficult. However the real difficulty is for gauge theories where gauge constraint makes it complicated. Extended Hilbert space approaches are probably useful in such situations just as entanglement entropy case, see for example, \cite{Ghosh:2015iwa, Soni:2015yga, Aoki:2015bsa, Aoki:2017ntc}.

\vspace{5mm}
\centerline{\bf{Acknowledgments}} 
\vspace{1mm}
\begin{acknowledgments}
Part of this work was presented by NI at the YITP workshop, ``Holography, Quantum Entanglement and Higher Spin Gravity II'', and RIKEN-Osaka-OIST Joint Workshop 2018 at OIST. 
We would like to thank the audience for helpful feedback, especially Javier Magan and Henry Maxfield. 
We would also like to thank Keun-Young Kim for stimulating discussions, and Javier Magan for useful comments on the manuscript.
The work of K.H. was supported 
in part by JSPS KAKENHI Grants No.~JP15H03658, 
No.~JP15K13483, and No.~JP17H06462. The work of 
N.I. was supported in part by JSPS KAKENHI Grant 
No.~JP25800143. S. S. is supported in part by the Grant-in-Aid for JSPS Research Fellow, Grant No.~JP16J01004.
\end{acknowledgments}


\end{document}